\documentclass[journal]{IEEEtran}
\newcommand{\bfs}{\boldsymbol}

\newcommand{\rl}[1]{#1} % Turn off highlighting
\include{asmath}
\usepackage{graphicx}
\usepackage{threeparttable}
\usepackage{cite}
\usepackage{subcaption}
\usepackage{color}
\usepackage{amsfonts}
\usepackage{threeparttable}
\usepackage{tabularx}
\usepackage{fixltx2e}
\usepackage{hhline}
\usepackage{amsmath,amssymb}
\usepackage{dblfloatfix}

\usepackage{todonotes}
\begin{document}
\bstctlcite{IEEEexample:BSTcontrol}
\title{Impact of 4D Channel Distribution on the Achievable Rates in Coherent Optical Communication Experiments}

\author{Tobias A. Eriksson, Tobias Fehenberger, Peter A. Andrekson, Magnus~Karlsson, Norbert Hanik, and Erik Agrell %<-this % stops a space

\thanks{Tobias~A.~Eriksson, Peter~A.~Andrekson, and Magnus~Karlsson are with the Photonics Laboratory, Department of Microtechnology and Nanoscience, Chalmers University of Technology, SE-412 96 Gothenburg, Sweden (e-mail: tobias.eriksson@chalmers.se)}
\thanks{Tobias Fehenberger and Norbert Hanik are with the Institute for Communications Engineering, Technical University of Munich, 80333 Munich, Germany}%
\thanks{Erik Agrell is with the Communication Systems Group, Department of Signals and Systems, Chalmers University of Technology, SE-412 96 Gothenburg, Sweden.}%
\thanks{This work was partially funded by the Swedish Research Council (VR).}
\thanks{Copyright (c) 2016 IEEE. Personal use of this material is permitted.  However, permission to use this material for any other purposes must be obtained from the IEEE by sending a request to pubs-permissions@ieee.org.}%
}

% The paper headers
%\markboth{PREPRINT}%
%{Eriksson \MakeLowercase{\textit{et al.}}: $K$-over-$L$ Multidimensional Position Modulation}

% make the title area
\maketitle

\begin{abstract}
We experimentally investigate mutual information and generalized mutual information for coherent optical transmission systems. The impact of the assumed channel distribution on the achievable rate is investigated for distributions in up to four dimensions. Single channel and wavelength division multiplexing (WDM) transmission over transmission links with and without inline dispersion compensation are studied. We show that for conventional WDM systems without inline dispersion compensation, \rl{a} circularly symmetric complex Gaussian distribution is a good approximation of the channel. For other channels, such as with inline dispersion compensation, this is no longer true and gains in the achievable information rate are obtained by considering more sophisticated four-dimensional (4D) distributions. We also show that for nonlinear channels, gains in the achievable information rate can also be achieved by estimating the mean values of the received constellation in four dimensions. The highest gain for such channels is seen for a 4D correlated Gaussian distribution.
\end{abstract}

\begin{IEEEkeywords}
Channel models, Digital communication, Fiber nonlinear optics, Mutual Information, Optical fiber communication.
\end{IEEEkeywords}

\section{Introduction}
\IEEEPARstart{C}{oherent} optical communication systems have largely been enabled by the use of digital signal processing (DSP) \cite{SavoryDSP} which is used to mitigate signal distortions such as laser phase drifts and polarization drifts. This has eased the use of higher order modulation formats that make use of both the phase and the amplitude of the optical field. The driving force to increase the modulation order is the ever-increasing demand for increased data rates in today's communication systems \cite{EsiambreCapacity}, as a higher order modulation format enables transmission with a higher spectral efficiency (SE).

One of the key technologies in present communication systems is forward error correction (FEC) coding. The use of FEC can increase the sensitivity of a communication system significantly at the cost of lower spectral efficiency, due to the overhead from the code, and increased complexity in the encoding and decoding circuitry. Without the use of FEC, many higher-order modulation formats such as polarization-multiplexed 16-ary quadrature amplitude modulation (PM-16QAM) cannot be transmitted \rl{over any significant distance} before the detected data can no longer be considered "error-free" (typically defined as bit error rate (BER) $<10^{-15}$) \cite{ZhangPM16QAM}. However, using advanced FEC, PM-16QAM can be transmitted over transoceanic distances \cite{Cai16QAM}. The use of advanced FEC in combination with PM quadrature phase shift keying (PM-QPSK) has enabled record SE-transmission distance products \cite{QianRecordSEDistance,IgarashiExabit}.

Up until recent years, the FEC codes used in \rl{fiber-optic} transmission systems were typically Reed-Solomon (RS) \cite{ITUreedSolomon} or Bose-Chaudhuri-Hocquenghem (BCH) codes \cite{ITUbCH}, decoded using hard-decision (HD) algorithms. This means that the received constellation is demapped into bits before information is passed to the decoder. In recent years, the FEC codes considered for \rl{fiber-optic} systems are based on soft-decision (SD) decoding, which means that the soft information from the received constellation after the receiver DSP is sent to the decoder. Examples of such codes are low-density parity check codes (LDPC) \cite{GallagerLDPC}, polar codes \cite{ArikanPolarCodes}, and turbo codes \cite{BerrouTPC}. These types of codes can achieve significantly higher sensitivity compared to the HD codes at the cost of higher complexity. It should be noted that in some cases, concatenated codes with an inner SD code and an outer HD code are considered, which is practical if the inner code has an error-floor above the error-free BER limit.

\begin{figure*}[!h]
\centering
\includegraphics[width=1\textwidth]{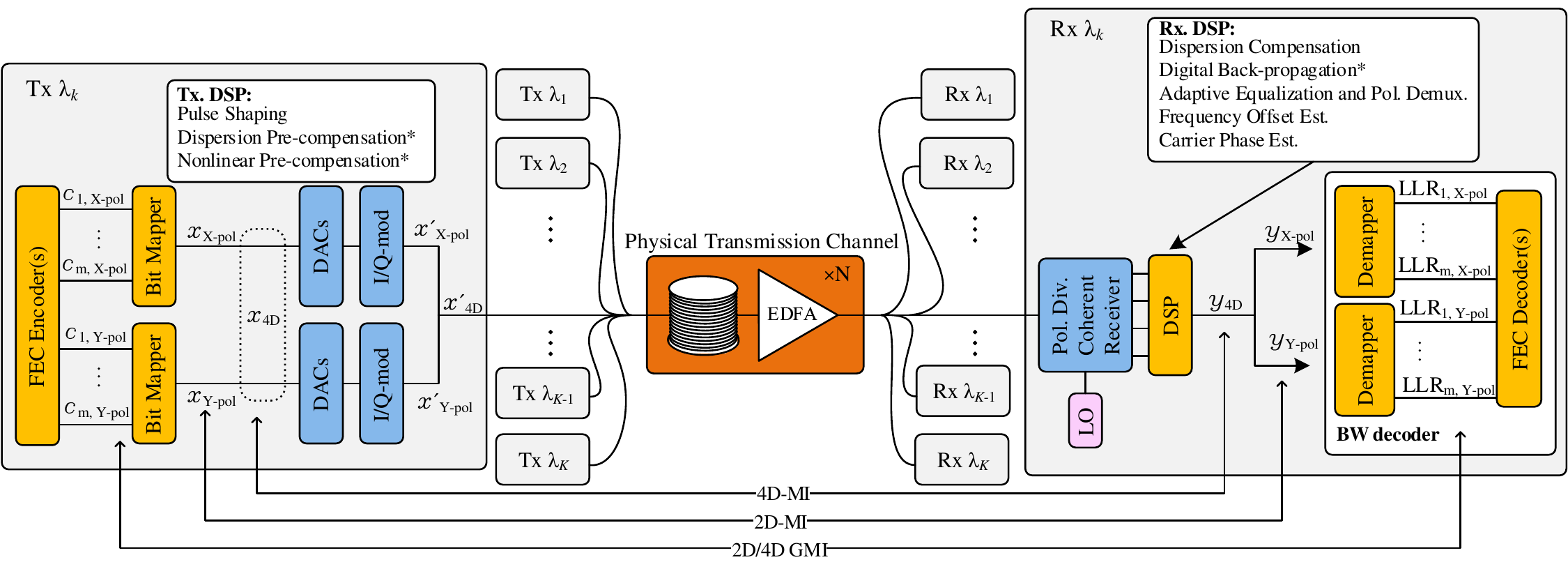}
\caption{Schematic of a typical fiber optical communication system. Items listed under Tx. DSP and Rx. DSP functions that are marked with * are not implemented in this work.}
\label{Fig_systemDescription}
\end{figure*}

\rl{For HD coding schemes, if the channel seen by the FEC can be modeled as a binary symmetric channel, the achievable rate after decoding can be fully determined by the pre-FEC BER \cite{MizuochiFEC}.} In other words, estimating the BER from the received constellation gives a good estimate of the post-FEC BER and is indeed what is done in most \rl{fiber-optic} transmission experiments. The main reason that the FEC decoder is not implemented in experiments is that the number of samples needed for good statistics at a post-FEC BER of $10^{-15}$ is infeasible with offline processing. For SD decoders on the other hand, there exists no such relation between the pre-FEC BER and the achievable rate. This can be explained by the fact that an SD decoder does not work on bits as input. For optical communication systems with SD FEC, it has been shown that mutual information (MI) is a more reliable measure than the pre-FEC BER \cite{LevelMI}. In \cite{AlvaradoReplacingSoftFEC}, it is shown that for an LDPC code and a turbo-product code (TPC) with  bit-wise decoders, the generalized mutual information (GMI) gives a good estimate of the post-FEC BER independently of the modulation format and optical launch power for a link without inline dispersion compensation.

Most of the previous work on mutual information as a figure of merit in \rl{fiber-optic} communication systems has been carried out theoretically or with simulations \cite{DjordjevicMI, EssiambreMI, Secondini2013JLT, AlvaradoReplacingSoftFEC, FehenbergerAchivRates, AgrellCapacityNonlinear, KramerCapacity}, or in experiments where the output statistics of the channel is considered as memoryless with additive Gaussian noise statistics with the same variance in all dimensions \cite{LevelMI,MillarMI,MaherMIdbp,maher2015modulation}.

However, the \rl{fiber-optic} channel is inherently nonlinear, which when approximated as memoryless white Gaussian noise means that information is lost. This has been discussed in several publications \cite{EssiambreMI, Secondini2013JLT, AgrellCapacityNonlinear, MarsellaMLSD, KramerCapacity}. In \cite{ErikssonMI}, we study the impact of different four-dimensional (4D) assumption of the memoryless channel distribution on the achievable information rate. We show that for single-channel transmission of 14~Gbaud PM-16QAM in a transmission system with inline dispersion compensation with high launch powers, \rl{significant gains in the achievable information rate can be achieved by employing receiver-side models that assume 4D channel distributions.}

%significant gains in the achievable information rate can be achieved by 4D assumptions of the channel distribution.
%%\te{It is interesting to note that the only upper bound on the MI that exists today is \cite{KramerCapacity}, where an upper bound on a channel that can be modeled using the split-step Fourier method is derived. The reason why it is hard to give upper bounds for fiber-optical transmission channels, is the complicated interactions between the nonlinear distortions, WDM-channels, amplified spontaneous emission (ASE) noise and chromatic dispersion. (Remove completely?)}

In this work we experimentally investigate different estimates of the achievable rate using both MI and GMI. We compare different \rl{models} of the channel distribution and its impact on the achievable rate for different transmission scenarios. We show that for most realistic scenarios, circularly symmetric Gaussian noise statistics is a good assumption. \rl{We also show that for some specific links, a higher achievable rate can be achieved by using more sophisticated channel models, based on 4D distributions, in the decoder.}

\section{Achievable Information Rates}
Fig.~\ref{Fig_systemDescription} shows a schematic of a typical long-haul point-to-point fiber optical communication system for which different estimates of the achievable information rate will \rl{be} derived in this section.

\subsection{Mutual Information}
We follow the approach of \cite{Arnold,Secondini2013JLT,FehenbergerAchivRates} to evaluate a lower bound estimate on the symbolwise mutual information (MI) from real $d$-dimensional  symbols that are obtained after transmission and DSP. Let the channel input $\bfs{X}$ be a $d$-dimensional random variable that is drawn from a constellation $\mathcal{X}=\{\mathbf s_1,\ldots,\mathbf s_M\}$ with cardinality $|\mathcal{X}|=M=2^m$ with identical probability. The channel output $\bfs{Y}$ denotes a random variable that is dependent on $\bfs{X}$ and the channel. The MI between  $\bfs{X}$ and $\bfs{Y}$ is given as
\begin{equation}
  I(\bfs{X};\bfs{Y}) = \text{E}_{\bfs{XY}} \!\! \left[ \log_2 \frac{p_{\bfs{Y}|\bfs{X}}(\bfs{Y}|\bfs{X})}{p_{\bfs{Y}}(\bfs{Y})} \right],
  \label{eq:MImemory}
\end{equation}
where $p_{\bfs{Y}|\bfs{X}}$ is the \rl{conditional} memoryless channel transition rl{distribution} and $p_{\bfs{Y}}$ the channel output density. The MI gives the highest rate for a specific channel and input distribution at which reliable communication is possible and cannot exceed $m$ bits per channel use. In other words, below the MI there exist codes with the possibility of a post-FEC BER that approaches~0. As indicated in Fig.~\ref{Fig_systemDescription}, the MI is calculated on a symbol level, i.e., the bit-to-symbol mapping does not influence the MI.

Note that \eqref{eq:MImemory} gives the MI for a memoryless channel. In reality, a \rl{fiber-optic} channel exhibits memory, which makes $I(\bfs{X};\bfs{Y})$ a lower bound on the MI calculated with the channel input and output as sequences  \cite[Sec.~III-F]{Essiambre2010}\cite{Verdu}. We also note that considering a finite memory due to nonlinear effects of the \rl{fiber-optic} channel can have a large effect on the achievable information rate \cite{AgrellCapacityNonlinear}. Since most of the linear memory introduced in the optical fiber channel, such as dispersion and inter-symbol interference, is compensated by the DSP, we restrict our analysis to the memoryless MI for the remainder of this work. The experimental results in Section~\ref{section:expResults} nevertheless \rl{apply} to the true \rl{fiber-optic} channel with residual memory after DSP.

Using the weak law of large numbers, we can estimate $I(\bfs{X};\bfs{Y})$ of \eqref{eq:MImemory} via Monte-Carlo integration from $N$ input-output pairs $(\bfs{x}_i,\bfs{y}_i)$ \cite{Arnold} as
\begin{equation}
  \frac{1}{N} \sum_{i=1}^{N} \log_2 \frac{p_{\bfs{Y}|\bfs{X}} (\bfs{y}_i|\bfs{x}_i)}{p_{\bfs{Y}}(\bfs{y}_i)} \xrightarrow{p} I(\bfs{X};\bfs{Y}),
  \label{eq:knownChannelMonteCarlo}
\end{equation}
where $\xrightarrow{p}$ denotes convergence in probability. Eq.~\eqref{eq:knownChannelMonteCarlo} implies that if there is an analytical description of the channel over which we have transmitted $N$ symbols, an estimate of the MI is obtained whose accuracy increases with $N$.

In fiber optical communications, however, the channel transition \rl{distribution} $p_{\bfs{Y}|\bfs{X}}$ is not known, i.e., Eq.~\eqref{eq:knownChannelMonteCarlo} cannot be evaluated directly. Following \cite{Arnold}, it can be shown that a lower bound on $I(\bfs{X};\bfs{Y})$ is achieved by using mismatched decoding. The samples at the channel output, after fiber transmission, are evaluated as if they were transmitted over an auxiliary channel with transition \rl{distribution} $q_{\bfs{Y}|\bfs{X}}$ instead of the true yet unknown channel $p_{\bfs{Y}|\bfs{X}}$. Note that $q_{\bfs{Y}|\bfs{X}}$ has the same input and output alphabet as $p_{\bfs{Y}|\bfs{X}}$, i.e., $q_{\bfs{Y}}(\bfs{y})= \sum_{\bfs{x} \in \mathcal{X}} q_{\bfs{Y}|\bfs{X}}(\bfs{y}|\bfs{x}) \cdot P_{\bfs{X}}(\bfs{x})$. Since MI is an achievable rate, a lower bound on the MI will also be an achievable rate. We denote this lower bound as ${R}$ and define it as
\begin{align}\label{eq:mismatcheddecoderMI}
	I(\bfs{X};\bfs{Y}) &= \text{E}_{\bfs{XY}} \left[ \log_2 \frac{p_{\bfs{Y}|\bfs{X}}(\bfs{Y}|\bfs{X})}{p_{\bfs{Y}}(\bfs{Y})} \right] \nonumber \\
	& \geq \text{E}_{\bfs{XY}} \left[ \log_2 \frac{q_{\bfs{Y}|\bfs{X}}(\bfs{Y}|\bfs{X})}{q_{\bfs{Y}}(\bfs{Y})} \right] \triangleq {R}.
\end{align}
Throughout this paper, we define $R$ in units of bit per 4D-symbol (bit/4D-sym). It is apparent that the better $q_{\bfs{Y}|\bfs{X}}$ resembles $p_{\bfs{Y}|\bfs{X}}$, the tighter the bound in \eqref{eq:mismatcheddecoderMI} will be, and a higher achievable rate is obtained.  Although we do not obtain the true MI of the channel, the mismatched decoder approach gives a practical achievable rate since a decoder would also have to assume a channel.
Using the auxiliary channel $q_{\bfs{Y}|\bfs{X}}$, a lower bound on MI is estimated in the same fashion as in \eqref{eq:knownChannelMonteCarlo},
\begin{equation}
 \frac{1}{N} \sum_{i=1}^{N} \log_2 \frac{q_{\bfs{Y}|\bfs{X}} (\bfs{y}_i|\bfs{x}_i)}{q_{\bfs{Y}}(\bfs{y}_i)} \xrightarrow{p} {R},
 \label{eq:unknownChannelMonteCarlo}
\end{equation}
where $\bfs{x}_i$ and  $\bfs{y}_i$, as in \eqref{eq:knownChannelMonteCarlo}, are obtained experimentally from the true channel $p_{\bfs{Y}|\bfs{X}}$.

In this paper, we assume $q_{\bfs{Y}|\bfs{X}}$ to be $d$-dimensional Gaussian distributed,
\begin{equation} \label{eq:auxchannel}
q_{\bfs{Y}|\bfs{X}}(\bfs{y}|\bfs{s}_j)=
\frac{1}{(2\pi)^{\frac{d}{2}}|\bfs\Sigma_j|^{\frac{1}{2}}} e^{ -\frac{1}{2}(\bfs{y}-\bfs\mu_j)^\text{T}\bfs\Sigma_j^{-1}(\bfs{y}-\bfs\mu_j) }
\end{equation}
for $j = 1, \dots, M$ and $\bfs{s}_j$ denoting the $j$\textsuperscript{th} constellation point. Further, $\bfs{x}$ and $\bfs{y}$ are real $d$-dimensional column vectors and $|\bfs\Sigma_j|$ is the determinant of the covariance matrix $\bfs\Sigma_j$. We discuss in detail in Section~\ref{ssec:paramestimation} how the mean values $\bfs\mu_j$ and the covariances $\bfs\Sigma_j$ of \eqref{eq:auxchannel} are obtained. The choices for the auxiliary channel in this work are presented in Section~\ref{ssec:channelmodels}.

\subsection{Generalized Mutual Information}
In addition to MI, we also compare achievable rates using GMI as a figure of merit. GMI gives an achievable rate for the bit-wise (BW) decoder \cite{CaireBICM} and has been shown to be an accurate estimate of the post-FEC BER for a wide range of channels, modulation formats, and codes \cite{AlvaradoReplacingSoftFEC,Alvarado4DGMI}. It should be noted that, in the same way as for the MI estimates, this is a mismatched decoder approach. While it gives an practically achievable rate, it does not give the highest achievable rate for the BW decoder. However, for decoders applying iterative demapping and decoding \cite{DjordjevicLDPCcoded,BulowCodedModulation}, it is yet to be validated how accurately GMI represents the post-FEC performance.

Assuming uniformly distributed transmitted symbols, the GMI is estimated as \cite{AlvaradoReplacingSoftFEC}
\begin{equation} \label{eq:GMI}
\textrm{GMI} \approx m - \frac{1}{N} \sum_{k=1}^m\sum_{i=1}^{N} \log_2\left ( 1+e^{(-1)^{b_{k,i}} \textrm{LLR}_{k,i} }\right ),
\end{equation}
where $b_{k,i}$ is the transmitted bit sequences and $\textrm{LLR}_{k,i}$ is the log-likelihood \rl{ratio} with $k$ denoting the bit position and $i$ denoting the $i^{\text{th}}$ received symbol. In analogy to the MI computation in \eqref{eq:mismatcheddecoderMI}, the calculation of LLRs also requires an assumption on the underlying channel distribution. \rl{In addition to finding the best-possible matching of the auxiliary channel and  the true channel distribution, an optimization over a non-negative parameter is in principle required, as discussed in \cite[Sec.~III-C]{AlvaradoReplacingSoftFEC}. This optimization is omitted in this work, possibly resulting in a lower GMI estimate.}

%Note that the LLRs cannot be matched to the unknown optical channel and, in principle, an optimization over a non-negative parameter is required in \eqref{eq:GMI} \cite[Sec.~III-C]{AlvaradoReplacingSoftFEC}. This optimization is omitted in this work, possibly resulting in a lower GMI estimate.

For the general case of a $d$-dimensional Gaussian distribution as defined in \eqref{eq:auxchannel}, the LLRs are calculated as \cite{MartinezGMI,AlvaradoReplacingSoftFEC}

\begin{align} \label{Eq:4DLLRs}
\text{LLR}_{k,i} =
\log \frac{q_{\mathbf Y| B_k}{(\mathbf y_i|1)}}{q_{\mathbf Y| B_k}{(\mathbf y_i|0)}}   =
\log \frac{
		\sum\limits_{j: \mathbf{s}_j \in \mathcal{X}_1^k}  q_{\mathbf Y|\mathbf X}{(\mathbf{y}_i|\mathbf s_j)}
		}
		{
		\sum\limits_{j: \mathbf{s}_j \in \mathcal{X}_0^k}	 q_{\mathbf Y|\mathbf X}{(\mathbf{y}_i|\mathbf s_j)}}  \nonumber \\
= \log \frac{
		 \sum\limits_{j: \mathbf{s}_j \in \mathcal{X}_1^k}  \frac{1}{\sqrt{(2\pi)^d |\boldsymbol\Sigma_j|}}
		\exp \! \left( \! -\frac{1}{2}(\mathbf{y}_i-\boldsymbol{s}_j)^\mathrm{T}\boldsymbol\Sigma_j^{-1}(\mathbf{y}_i-\boldsymbol{s}_j) \right)
		}
		{
		\sum\limits_{j: \mathbf{s}_j \in \mathcal{X}_0^k}  \frac{1}{\sqrt{(2\pi)^d |\boldsymbol\Sigma_j|}}
		\exp \! \left( \! -\frac{1}{2}(\mathbf{y}_i-\boldsymbol{s}_j)^\mathrm{T}\boldsymbol\Sigma_j^{-1}(\mathbf{y}_i-\boldsymbol{s}_j) \right)
		},
\end{align}
where $q_{\bfs{Y}|B_k}$ denotes the probability density function of the auxiliary channel conditioned on transmitting the $k$\textsuperscript{th} bit $B_k$ of $\bfs{X}$. Further, $\mathcal{X}_0^k$ and $\mathcal{X}_1^k$ are the sets of constellations points where the $k$\textsuperscript{th} bit equals 0 and 1, respectively. It is important to note that, as indicated in Fig.~\ref{Fig_systemDescription}, the GMI is dependent on the bit-to-symbol mapping. Throughout this paper, we use Gray-coded constellations in two dimensions.

\subsection{Parameter Estimation}\label{ssec:paramestimation}
We can see from \eqref{eq:auxchannel} that the dimensionality and the choice of $\bfs\mu_j$ and $\bfs\Sigma_j$ define the Gaussian auxiliary channel and thus, both MI and GMI. In this work, we differentiate between \textit{static} and \textit{adaptive} mean values. \rl{Note for the adaptive case, the mean values are estimated for a batch of received symbols but kept constant over this batch. This should be differentiated from a case where the mean values are tracked over time, which is not in the scope of this paper.} The covariances are taken as either independent and identically distributed Gaussian (iidG) or correlated Gaussian (CG).

When static mean values are used, $\bfs\mu_j$ is one of the $M$ values of the input alphabet $\mathcal{X}$, i.e., $\bfs\mu_j=\bfs{s}_j,~\forall j$. For adaptive mean values, we use the conditional sample mean $\bfs\mu_j$ for the $j$\textsuperscript{th} mean of the multivariate normal distribution,
\begin{equation} \label{eq:condSampleMean}
\bfs\mu_j = \frac{1}{|\mathcal{I}_j|} \sum_{i \in \mathcal{I}_j} \bfs{y}_i,\end{equation}
where the index set $\mathcal{I}_j$ denotes the indices of all $\bfs{y}_i$'s that correspond to a sent constellation point $\bfs{x}_i$ as
$\mathcal{I}_j = \{ i \in \{1,\ldots, N \}: x_i = s_j \}$.

For iid Gaussian auxiliary channels, $\bfs\Sigma_j$ is a $d \times d$ identity matrix $I_d$ multiplied with the average one-dimensional noise variance,
\begin{equation} \label{eq:condSampleVariance}
\sigma^2_\text{1D} = \frac{1}{d} \sum_{l=1}^{d} \frac{1}{N-1} \sum_{i=1}^{N} (\bfs{y}_{i,l}-\bfs\mu_{j,l})^2,\end{equation}
where the index $l$ refers to the $l$\textsuperscript{th} dimension of a $d$-dimensional vector. Note that \rl{$\sigma^2_\text{1D}$} is identical for all $j$'s, i.e., $\bfs\Sigma_j=\sigma^2_\text{1D}\cdot I_d, ~ \forall j$.
In the case of correlated Gaussian auxiliary channels, $M$ $\bfs\Sigma_j$'s are calculated, one for each constellation point.  The sample covariance conditioned on the $j$\textsuperscript{th} constellation point being sent is then estimated from the received samples $\bfs{y}$ as
\begin{equation} \label{eq:condSampleCovariance}
\bfs\Sigma_j = \frac{1}{|\mathcal{I}_j|-1} \sum_{i \in \mathcal{I}_j} (\bfs{y}_i-\bfs\mu_j)^\text{T} (\bfs{y}_i-\bfs\mu_j).\end{equation}
The estimates of \eqref{eq:condSampleMean} and \eqref{eq:condSampleCovariance} have the same dimensionality $d$ as the auxiliary channel in \eqref{eq:auxchannel}. \rl{It is important to note that this conditional sample covariance means that the noise is not additive since $\bfs\Sigma_j$ varies with the transmitted constellation point $\bfs{s}_j$. In other words, the channel noise statistic is conditionally Gaussian.}

Throughout this paper, we randomly choose $N$ samples at the output of the DSP to estimate the parameters of $q_{\bfs{Y}|\bfs{X}}$ and $N$ different samples to calculate the achievable rate. \rl{For each estimated achievable rate, we use four different experimental batches, from each we take 200000 samples. The final achievable rate is an average over these four batches.} This double Monte Carlo approach ensures that we do not overestimate the achievable rate by estimating secondary parameters, i.e., the covariances and the mean values, and our figure of merit, i.e, the achievable rate, from the same sequence.

\begin{figure*}[ht]
\centering
        \begin{subfigure}[b]{0.5\columnwidth}
                \includegraphics[width=1\textwidth]{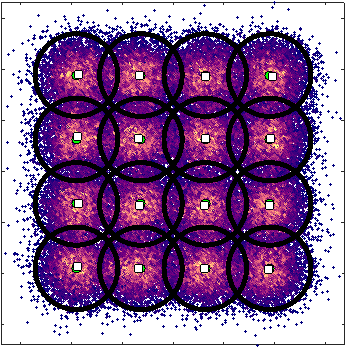}
                 \caption{2D-iidG : WDM wo. ILDC}
                 \label{fig:a}
        \end{subfigure}
        \begin{subfigure}[b]{0.5\columnwidth}
                \includegraphics[width=1\textwidth]{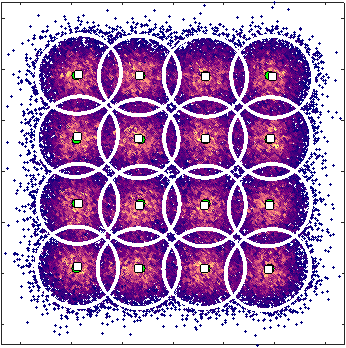}
                \caption{2D-CG : WDM wo. ILDC}
                \label{fig:b}
        \end{subfigure}
        \begin{subfigure}[b]{0.5\columnwidth}
                \includegraphics[width=1\textwidth]{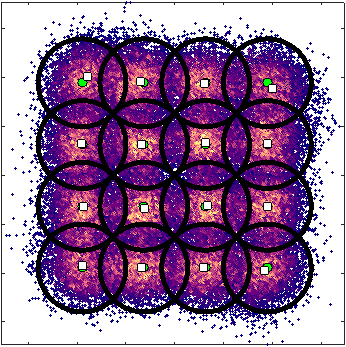}
                \caption{2D-iidG : SC w. ILDC}
                \label{fig:c}
        \end{subfigure}
        \begin{subfigure}[b]{0.5\columnwidth}
                \includegraphics[width=1\textwidth]{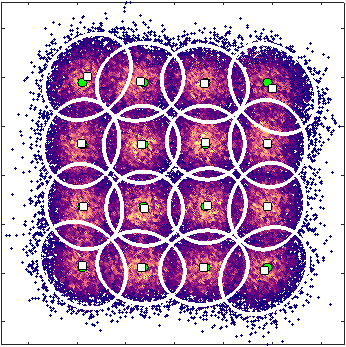}
                \caption{2D-CG : SC w. ILDC}
                \label{fig:d}
        \end{subfigure}
\caption{Measured constellations at $R \sim$ 7~bit/4D-sym illustrating the difference between 2D-CG and 2D-iiDG estimates showing (a) 2D-iidG and (b) 2D-CG for WDM transmission of 20~Gbaud PM-16QAM without inline dispersion compensation (ILDC). Further shown are (c) 2D-iidG (d) 2D-CG for single channel (SC) transmission of 20~Gbaud PM-16QAM with inline dispersion compensation. Circles and ellipses indicate the 90\% confidence interval of the variance for iidG (black) and the covariance for CG (white) estimates, respectively. The green \rl{circular markers} shows the transmitted constellation which is also what is used for fixed mean values. White \rl{square markers} show mean values estimates as per \eqref{eq:condSampleMean}.}
\label{fig:distributionPlots}
\end{figure*}

\begin{table}[!t]
 \renewcommand{\arraystretch}{1.7}
  \begin{threeparttable}[b]
   \caption{\rl{Models of the distribution of  $q_{\bfs{Y}|\bfs{X}}$}}
   \label{Table1}
   \begin{tabularx}{\columnwidth}{l X c p{1.3cm}}
\hline
\hline
\textsc{Name} & \textsc{Description} & \textsc{Parameters} &\textsc{DoFs}\tnote{$\S$} \\
 \bf{1D-iidG}   & 1D iid Gaussian noise in each quadrature and adaptive mean values.\tnote{$\dagger$}  & {\begin{tabular}[t]{@{}c@{}c@{}}  $d=1$ \\[-.6em] $\bfs\mu_j$ as per \eqref{eq:condSampleMean} \\[-.6em] $\bfs\Sigma_j=\sigma^2_\text{1D}\cdot I_1$\end{tabular}} & {$(4\!+\!1)\!\cdot\!4\!=\!20$} \\
 % \bf{1D-iidG}   & 1D iid Gaussian noise in each quadrature.\tnote{$\dagger$}  & {\begin{tabular}[t]{@{}c@{}c@{}}  $d=1$ \\[-.6em] $\bfs\mu_j$ as per \eqref{eq:condSampleMean} \\[-.6em] $\bfs\Sigma_j=\Sigma, ~ \forall j$\end{tabular}} & 20 \\
 \bf{2D-iidG}   & 2D iid Gaussian noise in each polarization with static mean values.\tnote{$\ddagger$} & {\begin{tabular}[t]{@{}c@{}c@{}}  $d=2$ \\[-.6em] $\bfs\mu_j=\bfs{s}_j$\\[-.6em] $\bfs\Sigma_j=\sigma^2_\text{1D}\cdot I_2$\end{tabular}} & {$(0\!+\!1)\!\cdot\!2\!= \quad 2$} \\
 % \bf{2D-iidG}   & \tf{Reference case:} 2D iid Gaussian noise in each polarization with static mean values.\tnote{$\ddagger$} & {\begin{tabular}[t]{@{}c@{}c@{}}  $d=2$ \\[-.6em] $\bfs\mu_j=\bfs{s}_j$\\[-.6em] $\bfs\Sigma_j=\Sigma, ~ \forall j$\end{tabular}} & 2 \\
 \bf{2D-CG} & 2D correlated Gaussian (CG) noise and adaptive mean values.\tnote{$\ddagger$} & {\begin{tabular}[t]{@{}c@{}c@{}}  $d=2$ \\[-.6em] $\bfs\mu_j$ as per \eqref{eq:condSampleMean} \\[-.6em] $\bfs\Sigma_j$ as per \eqref{eq:condSampleCovariance}\end{tabular}} & {$(16\!\cdot\!2\!+\!16\!\cdot\!3)\!\cdot\!2\!=\!160$}\\
 % \bf{2D-CG} & 2D correlated Gaussian (CG) noise.\tnote{$\ddagger$} & {\begin{tabular}[t]{@{}c@{}c@{}}  $d=2$ \\[-.6em] $\bfs\mu_j$ as per \eqref{eq:condSampleMean} \\[-.6em] $\bfs\Sigma_j$ as per \eqref{eq:condSampleCovariance}\end{tabular}} & 160\\
 \bf{4D-iidG}   & 4D iid Gaussian noise and adaptive mean values. & {\begin{tabular}[t]{@{}c@{}c@{}}  $d=4$ \\[-.6em] $\bfs\mu_j$ as per \eqref{eq:condSampleMean} \\[-.6em] $\bfs\Sigma_j=\sigma^2_\text{1D}\cdot I_4$\end{tabular}} & {$(16^2\!\cdot\!4\!+\!1\!)\!=\!1025$} \\
 % \bf{4D-iidG}   & 4D iid Gaussian noise. & {\begin{tabular}[t]{@{}c@{}c@{}}  $d=4$ \\[-.6em] $\bfs\mu_j$ as per \eqref{eq:condSampleMean} \\[-.6em] $\bfs\Sigma_j=\Sigma, ~ \forall j$\end{tabular}} & 1025 \\
 \bf{4D-CG} & 4D CG noise and adaptive mean values. & {\begin{tabular}[t]{@{}c@{}c@{}}  $d=4$ \\[-.6em] $\bfs\mu_j$ as per \eqref{eq:condSampleMean} \\[-.6em] $\bfs\Sigma_j$ as per \eqref{eq:condSampleCovariance}\end{tabular}}  & {$(16^2\!\cdot\!4\!+\!16^2\!\cdot\!10\!)\!=\!3584$}\\
 % \bf{4D-CG} & 4D CG noise. & {\begin{tabular}[t]{@{}c@{}c@{}}  $d=4$ \\[-.6em] $\bfs\mu_j$ as per \eqref{eq:condSampleMean} \\[-.6em] $\bfs\Sigma_j$ as per \eqref{eq:condSampleCovariance}\end{tabular}}  & 3328\\

\hline
 \end{tabularx}
	\begin{tablenotes}
    \item [$\S$] The calculated number is the total for all four dimensions of PM-16QAM.
    \item [$\dagger$] The total $R$ is the sum of the $R$ in all four quadratures.
    \item [$\ddagger$] The total $R$ is the sum of the $R$ in both polarizations.
   \end{tablenotes}
\end{threeparttable}
\end{table}

\begin{figure*}[hb]
\centering
\includegraphics[width=1\textwidth]{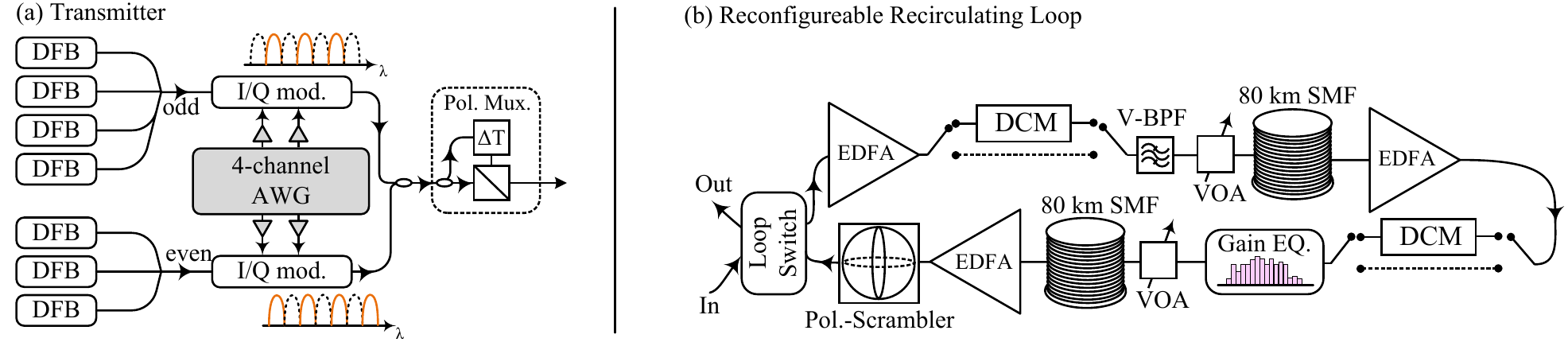}
\caption{(a) Schematics of the transmitter (b) Recirculating loop with 80~km spans. Abbreviations are explained in the text.}
\label{Fig_Tx_and_loop}
\end{figure*}

\subsection{Channel Models} \label{ssec:channelmodels}
\rl{In this work we consider five different \rl{models} of $q_{\bfs{Y}|\bfs{X}}$ using the Gaussian distributions that are given in Table~\ref{Table1}.} The 2D-iidG distribution is considered as a baseline since this is the most typical assumption made in the literature, i.e., all received constellation points are assumed to have the same variance in both dimensions and no adaptation of the mean values of the received constellation points is applied. We compare this to 2D-CG which assumes different covariances for each constellation point and also the center of each received constellation point is estimated. Since a polarization-multiplexed signal is indeed 4D, and the full 4D-field is required in the DSP, we investigate two 4D distributions. 4D-iidG assumes the same variance for all constellation points and in all four dimensions and 4D-CG estimates a 4D covariance matrix for each constellation point. Both of the 4D estimates applies adaptation of the center points of the received constellations in 4D. As PM-16QAM can be seen as a 1D modulation format, i.e. 4-ary pulse amplitude modulation in each dimension, we also include a 1D distribution where we assume the same variance for each constellation point and estimate the center of the 1D received constellation points. Shown in Fig.~\ref{fig:distributionPlots} are the received constellations in one polarization with $R \approx 7$ bit/4D-sym for two different experimental configurations, namely 20~Gbaud PM-16QAM in (a)--(b) WDM transmission without inline dispersion compensation and (c)--(d) single-channel transmission with inline dispersion compensation. Indicated in Fig.~\ref{fig:distributionPlots} is \rl{the 90~\% confidence interval of the variance, calculated from a Chi-square distribution with two free parameters}, as black circles for 2D-iidG and white circles for 2D-CG. Green \rl{circular markers} indicate the transmitted constellation and \rl{white square markers} the estimated mean values.

Also listed in Table~\ref{Table1} are the degrees of freedom (DoFs) for each distribution considered for the full 4D signal. The DoFs denote the number of parameters in $\bfs{\mu}_j$ and $\bfs{\Sigma}_j$ that are estimated for each auxiliary channel. The expression in the last column of Table~\ref{Table1} denotes the number of estimated mean values plus the number of covariances, multiplied with $4/d$ to get a 4D expression for the DoFs.
For 2D-iidG no mean values are estimated and a single variance is estimated per two dimensions, yielding total of 2 DoFs in four-dimensions. The 1D-iiG estimate with adaptive $\bfs{\mu}_j$'s has 4 mean values and one variance in each of the 4 four dimensions, resulting in 5 DoFs per 1D, or 20 DoFs in 4D. The remaining DoFs are calculated in analogy. Note that for the CG estimates, each $\bfs{\Sigma}_j$ has $d(d+1)/2$ DoFs because $\bfs{\Sigma}_j$ is a symmetric matrix. The number of DoFs is related to the complexity of the \rl{receiver}. However, the actual complexity will depend on the specific type of \rl{decoder} that is implemented, which is not in the scope of this paper. \rl{To reduce the DoFs, we note that it might be possible to exploit symmetries in $\bfs{\Sigma}_j$ and $\bfs{\mu}_j$. However, this is beyond the scope of this work.}

We also investigate GMI considering both 2D and 4D symbols, denoted as $\textrm{GMI}_{\textrm{2D}}$ and $\textrm{GMI}_{\textrm{4D}}$, which correspond to the cases of 2D-iidG and 4D-iidG in Table~\ref{Table1}, although both estimates use fixed mean values \rl{which is what is assumed in  most conventional bit-wise decoder structures \cite{Alvarado4DGMI}}. Thus, $\textrm{GMI}_{\textrm{2D}}$ has 2 DoFs and $\textrm{GMI}_{\textrm{4D}}$ has 1 DoF. \rl{However, in Section.~\ref{section:discussion}, we discuss the impact of using adaptive mean values for the GMI estimates}.

Note that the MI and GMI analysis applies to a discrete-time channel which in this case starts with the digital-to-analog converters (DACs), as shown in Fig.~\ref{Fig_systemDescription}. Choices that are made in the transmitter (Tx) on  pulse-shaping or dispersion pre-compensation will have an influence on the achievable rate since it actually changes the discrete-time channel. The same goes for the receiver (Rx) side, where the choices of algorithms will influence the achievable rate. An example that affects the results significantly is if any nonlinear compensation technique is used or not. More obvious perhaps is the impact from choices made on the actual transmission channel such as amplifier spacing, optical launch power, amplifier technology, and whether inline dispersion compensation is used or not.

\section{Experimental Setup}
The transmitter is shown in Fig.~\ref{Fig_Tx_and_loop}a. The electrical driving signals are generated using a 4-channel arbitrary waveform generator (AWG) running at 60~Gs/s, producing either 10~Gbaud or 20~Gbaud PM-16QAM signals with root-raised cosine (RRC) pulses using a roll-off factor of 0.5. The AWG drives two IQ modulators that are modulating in total 7 \rl{WDM channels} using distributed feedback lasers (DFBs) with $\sim$150~kHz linewidth as sources. For single-channel transmission, all lasers except the center channel are turned off. The IQ modulators are modulating the even and odd \rl{WDM channels} separately and the signals are \rl{decorrelated} in the AWG. After the IQ modulators, the optical signals are combined and sent to a polarization-multiplexing emulation stage based on split, delay, and recombination with orthogonal polarization states.

The optical signals are propagated over a recirculating loop shown in Fig.~\ref{Fig_Tx_and_loop}b. The loop consists of two spans of 80~km of conventional single-mode fiber (SMF) amplified by erbium-doped fiber amplifiers (EDFAs). Before the first span, a variable band-pass filter (V-BPF) is used to filter out excessive amplified spontaneous emission (ASE) noise. The bandwidth of the filter is varied depending on the type of signal transmitted. Before the second span, a programmable \rl{wavelength-selective} switch is used both as V-BPF and, when WDM-signals are transmitted, as a gain equalizer. Preceding both spans are variable optical attenuators (VOAs) that control the optical launch power. Also in the loop is a polarization scrambler (pol.-scrambler) that is synchronized to the round-trip time of the loop and a third EDFA compensates for the loss of the pol.-scrambler and the loop-switching components. The recirculating loop can be configured to have either inline dispersion compensation using \rl{dispersion-compensating} modules (DCMs) based on fiber-Bragg gratings \cite{proximon} or uncompensated by simply bypassing the DCMs. The DCMs are placed directly after the EDFAs that are preceding the fiber spans, as no extra EDFAs are then needed. It is important to note that the DCMs themselves inflict no nonlinear distortions \cite{TipsuwannakulFBG}.

\begin{figure}[t]
\centering
\includegraphics[width=0.8\columnwidth]{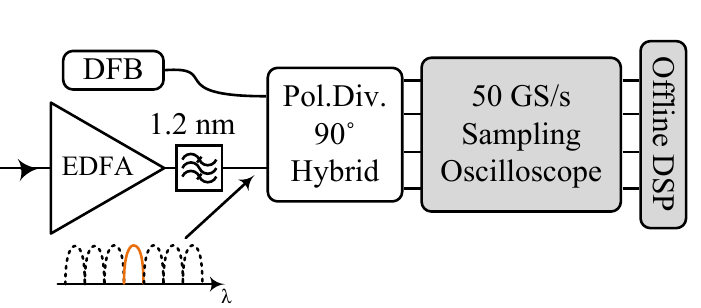}
\caption{Schematics of the coherent receiver.}
\label{Fig_rx}
\end{figure}

The center channel is detected using a coherent receiver based on an integrated polarization-diverse optical hybrid with balanced photo-detectors as depicted in Fig.~\ref{Fig_rx}. As local-oscillator (LO) a DFB laser of the same type as the transmitter lasers is used. The electrical signals are sampled using a 50~GS/s real-time oscilloscope with 33~GHz bandwidth and processed using off-line DSP.

\subsection{Digital Signal Processing}
The DSP starts with optical front-end compensation followed by resampling to 2 samples/symbol. If there is no inline dispersion compensation, electronic dispersion compensation (EDC) implemented in the frequency domain is applied. Polarization demultiplexing and adaptive equalization \rl{are} applied using four FIR filters in a butterfly structure. The filter taps are initially updated using the constant-modulus algorithm (CMA) for pre-convergence followed by decision-directed least mean square (DD-MLS) for final adaptation. The number of taps is either 17 for 20~Gbaud signals or 25 for 10~Gbaud. Frequency estimation based on the fast Fourier transform and carrier phase estimation (CPE) based on \rl{blind phase} search with 32 test angles \cite{PfauDSP} \rl{are} performed within the DD-LMS loop. The block length of the phase tracking is either 128 for 20~Gbaud signals or 256 for 10~Gbaud signals.

As mentioned previously, it should be noted that for the achievable rate estimates, the DSP is part of the channel and can affect the estimates. The adaptive equalizer and the phase tracking algorithms assume \rl{additive white Gaussian noise} statistics of the noise which, as is explained in the next section, is not the case for all of the systems investigated. Instead of using a blind receiver, a pilot-symbol based approach could be used \cite{ElschnerDataAided}. This type of receiver poses an interesting question of the trade-off between pilot overhead and FEC overhead. The most notable part of the DSP that influences the channel is the \rl{phase tracking}. For instance, we note that the block length of the CPE influences the nonlinear memory of the channel \cite{FehenbergerCPE}. Further, we limit ourself to not incorporating digital back propagation (DBP) in the DSP used in this study. It is known that DBP can greatly reduce the nonlinear distortions \cite{IpKahnDPB} at the cost of increased complexity. It has been shown that the achievable rate when DBP is used is dependent on the number of channels that are considered \cite{FehenbergerDBP}, and that its performance is dependent on frequency stabilization between the WDM channels \cite{TempranaDPBFreqStable}. Although interesting, we leave DBP for a future study.

\section{Experimental Results}\label{section:expResults}
In the following section, the achievable rates for different link configurations are presented. We investigate single channel transmission and WDM transmission of 10~Gbaud and 20~Gbaud PM-16QAM with and without inline dispersion compensation.

\begin{figure}[!t]
\centering
\includegraphics[width=1\columnwidth]{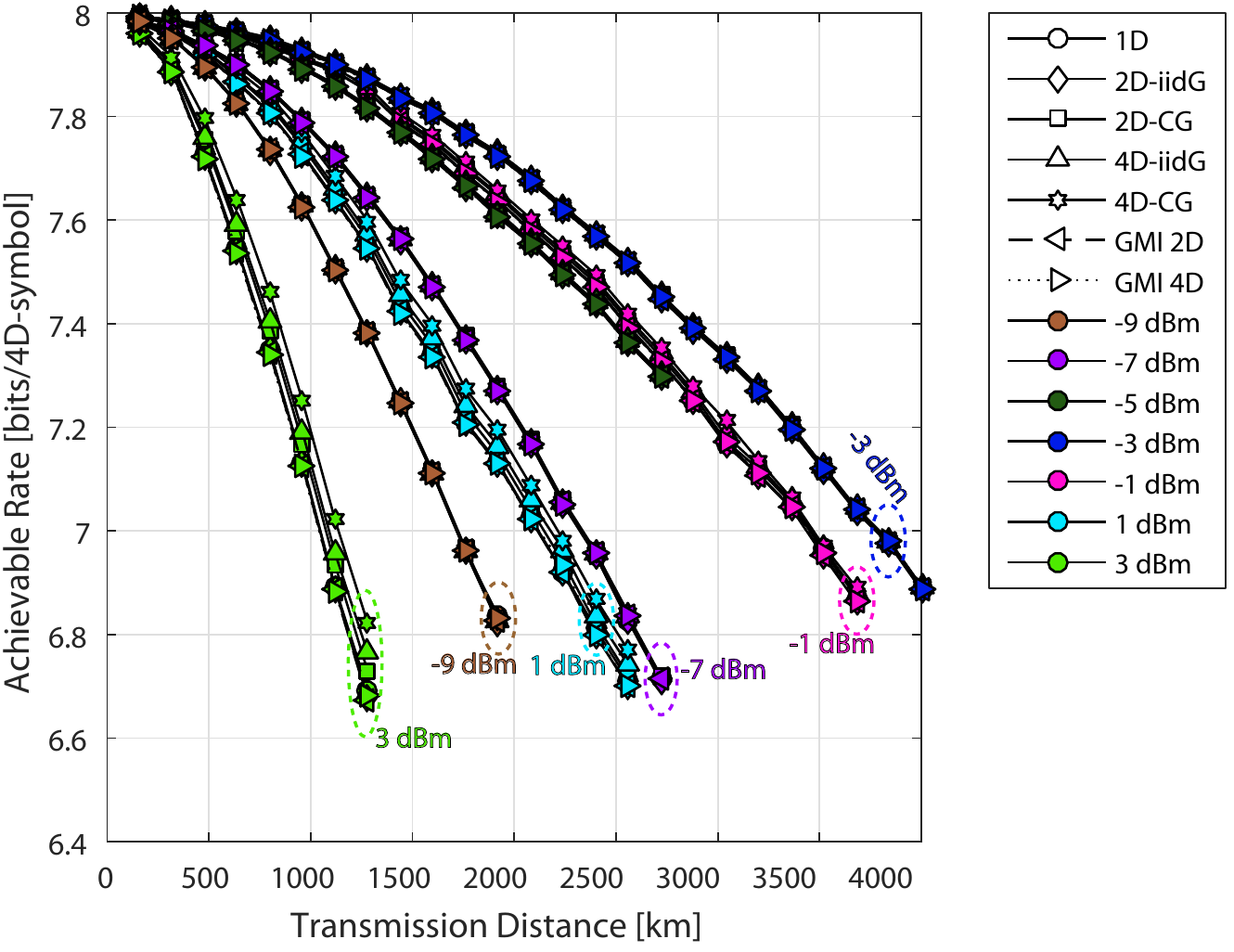}
\caption{WDM transmission of 20 Gbaud PM-16QAM with 30~GHz channel separation, without inline dispersion compensation.}
\label{Fig_PM16QAM_20Gbaud_30GHz_noDCM}
\end{figure}

\subsection{WDM transmission of 20~Gbaud PM-16QAM, no inline dispersion compensation}
In recent times, the majority of the fiber optical communication experiments reported are over transmission links without inline dispersion compensation. The main reason for this is that by accumulating the dispersion, in general the nonlinear impairments have less impact on the achievable transmission distances. Shown in Fig.~\ref{Fig_PM16QAM_20Gbaud_30GHz_noDCM} is the achievable rate using the different estimates as a function of transmission distance for different launch powers for WDM transmission of 20~Gbaud PM-16QAM with 30~GHz channel spacing. The main finding here is that the different assumptions on the distributions \rl{of} the achievable rate estimates have a negligible difference at the optimal launch power of $-$3~dBm. This is also true for lower launch power and slightly higher launch power. It is only in the extreme case of 1~dBm and 3~dBm launch power that significant difference between the estimates can be observed. At 1280~km and 3~dBm launch power 2D-GMI, 4D-GMI, 1D-iddG, and 2D-iidG \rl{give} roughly the same achievable rate. 2D-CG and 4D-iidG \rl{see} a small gain over 2D-iidG and the highest gain is seen by 4D-CG which has 0.13~bit/4D-symb higher achievable rate than the lowest cases. However, note that this is an unrealistically high launch power as the transmission distance is only 30~\% of that achieved with the optimal power. We may thus conclude that for \rl{WDM transmission} links without inline dispersion compensation, the memoryless noise statistics is iid Gaussian and a decoder working in 2D-iidG or even 1D-iidG should achieve as high rate as a receiver working with the more complex noise statistics. This is in good agreement with \cite{ErikssonMI} where we showed that the memoryless 2D-iidG \rl{channel distribution assumption} is valid for 28~Gbaud PM-QPSK and polarization-switched QPSK in (WDM) transmission without inline dispersion compensation.

\begin{figure}[!t]
\centering
\includegraphics[width=1\columnwidth]{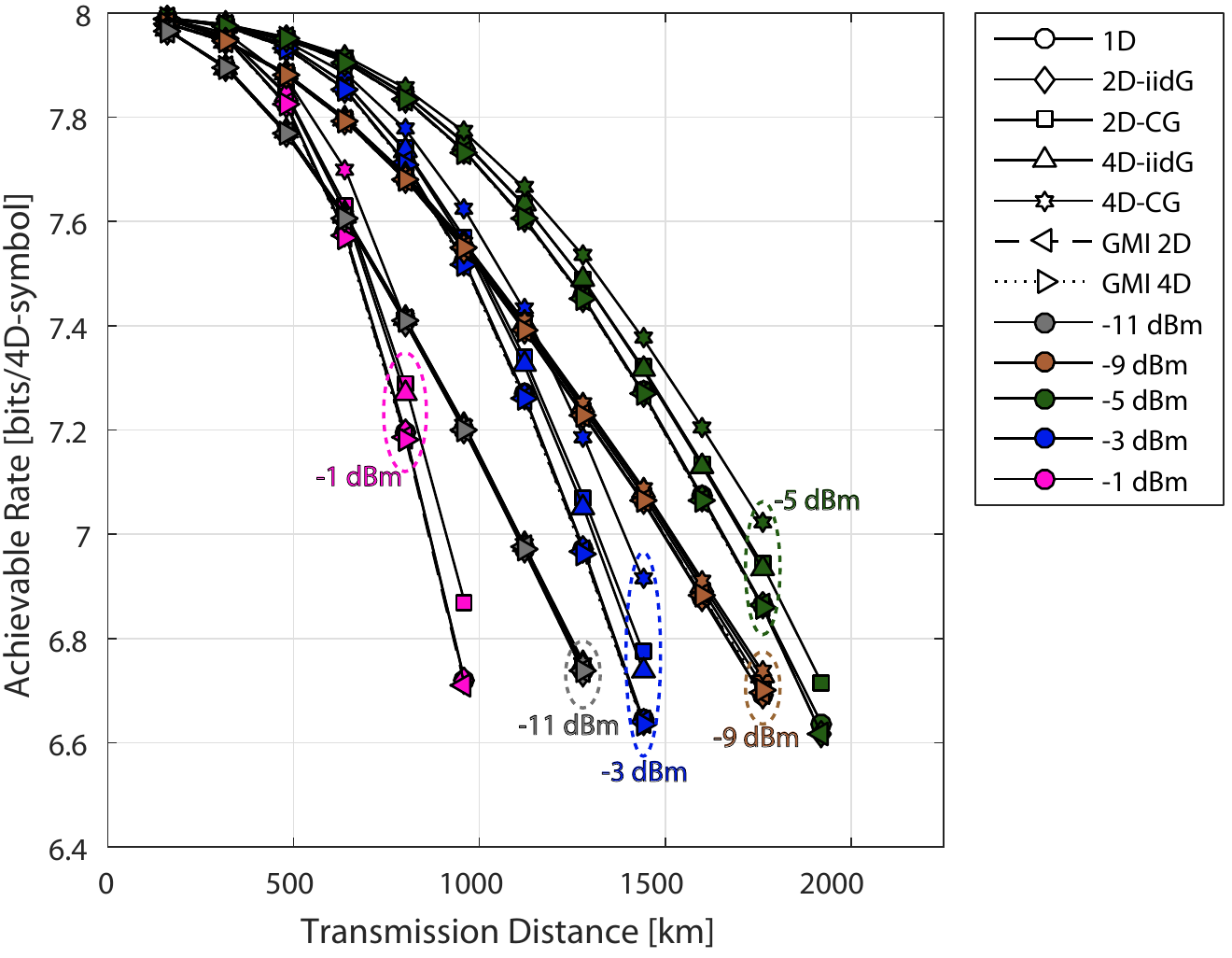}
\caption{WDM transmission of 20 Gbaud PM-16QAM with 30~GHz channel separation, with inline dispersion compensation. }
\label{Fig_PM16QAM_20Gbaud_30GHz_withDCM}
\end{figure}

\subsection{WDM transmission of 20~Gbaud PM-16QAM, with inline dispersion compensation}
Fig.~\ref{Fig_PM16QAM_20Gbaud_30GHz_withDCM} shows WDM transmission of 20~Gbaud PM-16QAM signals with a channel spacing of 30~GHz. For this case, inline dispersion compensation is used. \rl{Note that for clarity purpose, the results for $-$7~dBm have been removed as this curve is similar to that of $-$5~dBm.} We note that for the optimal launch power of $-$5dBm, the achievable transmission distance is severely reduced for this scenario compared to the previous case without inline dispersion compensation. Compared to the non-dispersion managed link with the same WDM signals, at an achievable rate of 7~bit/4D-sym, the transmission distance is roughly halved. We note that for very low launch powers, where nonlinear distortions can be neglected, there is no apparent difference between any of the estimates. However, for the optimal launch power there is a gain in the achievable rate by considering more sophisticated distributions. For this launch power, the difference between 2D-GMI, 4D-GMI, 1D-iidG, and 2D-iidG is insignificant. However,  4D-iidG and 2D-CG see a small gain, for instance at a distance of 1760~km the achievable rate is roughly 0.08~bit/4D-sym  higher than the previously mentioned estimates. The largest gain is seen by 4-CG which has a 0.16~bit/4D-sym higher achievable rate compared to the lowest estimates. As the launch power increases, the gain seen by the 4D-CG estimate increases.

\subsection{Single-channel transmission of 20~Gbaud PM-16QAM, without inline dispersion compensation}
\begin{figure}[!t]
\centering
\includegraphics[width=1\columnwidth]{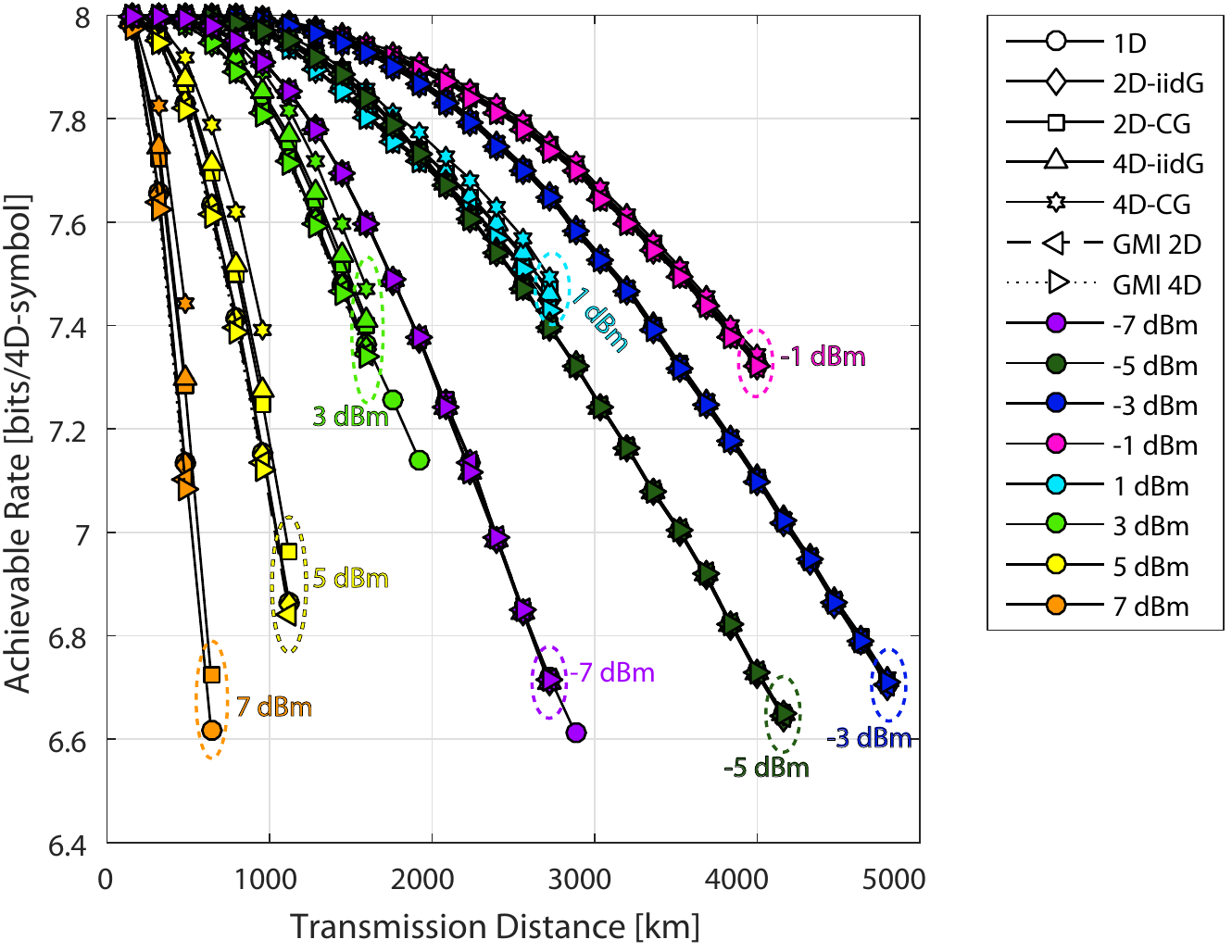}
\caption{Single channel transmission of 20 Gbaud PM-16QAM without inline dispersion compensation.}
\label{Fig_PM16QAM_20Gbaud_SC_noDCM}
\end{figure}

Shown in Fig.~\ref{Fig_PM16QAM_20Gbaud_SC_noDCM} are the achievable rates for all estimates as a function of transmission distance for different launch power for single-channel 20~Gbaud PM-16QAM without inline dispersion compensation. We notice that compared to the WDM case for the same link, the optimal launch power is increased by 2~dB. For low launch powers, there is no distinctive difference between the different estimates. In the same way as for the WDM case without inline dispersion compensation, for the highest launch powers, a difference can be observed. However, in this case the difference is significant already for launch powers that are close to the optimal and even for the optimal launch power a small gain can be observed for the 4D-CG estimate.

This means that for a single-channel system, even without inline dispersion compensation, the noise statistics are no longer iid Gaussian, especially for launch powers higher than the optimal. However, since at the optimal launch power the difference in achievable rate for the 4D-CG estimate and the 2D-iidG is less than 0.02 bit/4D-sym at 4000~km, it can be argued that even for this scenario, assuming iid Gaussian statistics of the channel transition \rl{distribution} is a reasonable trade-off between achievable rate and complexity.

\subsection{Single-channel transmission of 20~Gbaud PM-16QAM, with inline dispersion compensation}
\begin{figure}[!t]
\centering
\includegraphics[width=1\columnwidth]{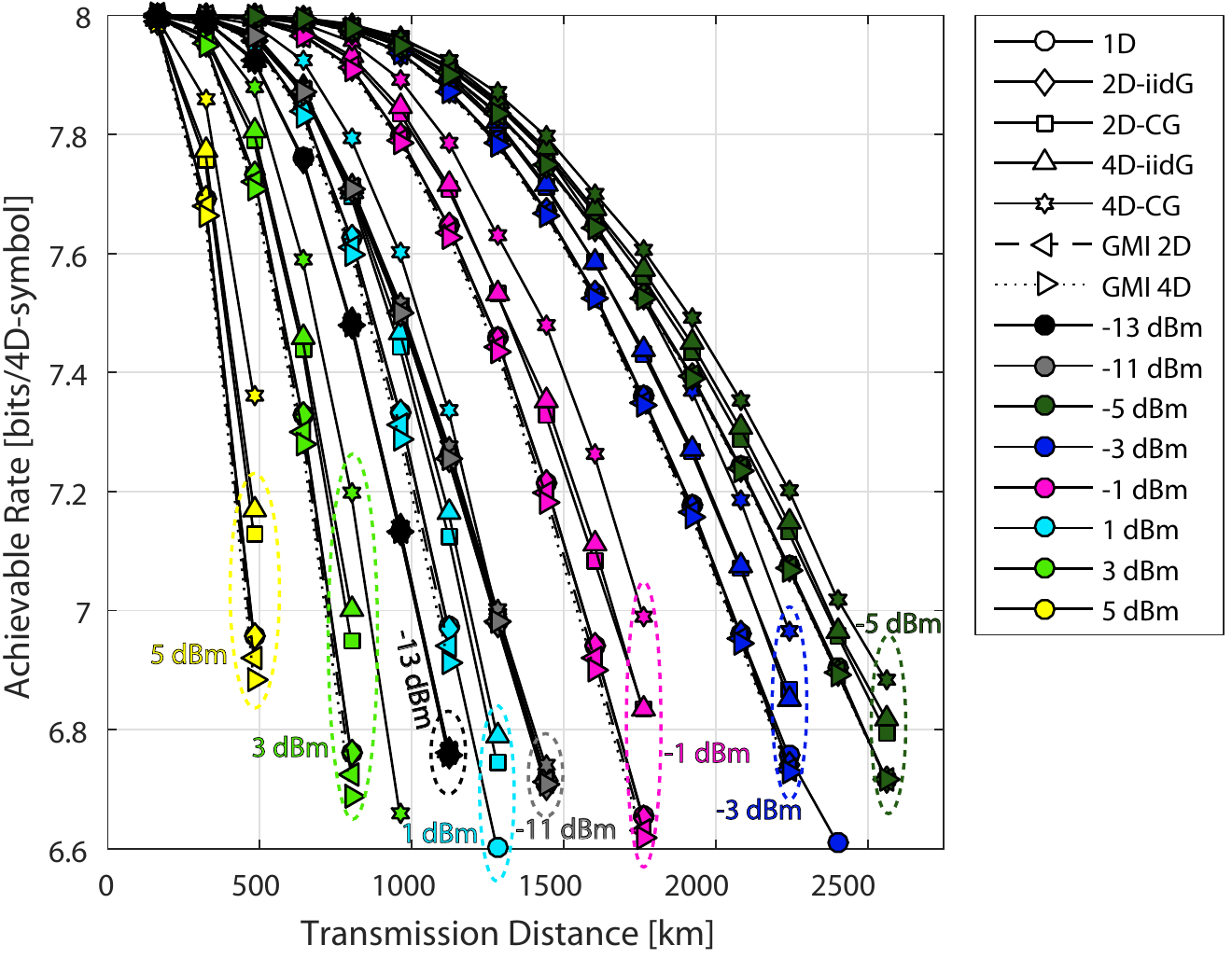}
\caption{Single channel transmission of 20 Gbaud PM-16QAM with inline dispersion compensation.} %Note that some launch powers have been omitted from the plot for clarity.}
\label{Fig_PM16QAM_20Gbaud_SC_withDCM}
\end{figure}

Presented in Fig.~\ref{Fig_PM16QAM_20Gbaud_SC_withDCM} is the achievable rate for single-channel 20~Gbaud PM-16QAM with inline dispersion compensation. \rl{Note that the results for $-$7~dBm launch power has been omitted for clarity.} The optimal launch power is $-$5~dBm, which is the same as for the WDM transmission over the same link. Compared to the single-channel transmission of 20~Gbaud PM-16QAM without inline dispersion compensation, we note that the optimal launch power is 4~dB lower. We also note that for this link, there is a big difference between the different estimates at the optimal launch power. At 2400~km, the 4D-CG estimate has a gain in achievable rate of roughly 0.13~bit/4D-sym over 2D-iidG. Further, 4D-iidG and 2D-CG see an intermediate gain of 0.07~bit/4D-sym and 0.06~bit/4D-sym, respectively. For this link, the noise statistics are not iid Gaussian except for low launch powers. For the optimal launch power, compared at an achievable rate of 7~bit/4D-symb, 5\% increased transmission distance can be achieved by assuming 4D-CG instead of 2D-iidG statistics.

\begin{figure}[!t]
\centering
\includegraphics[width=1\columnwidth]{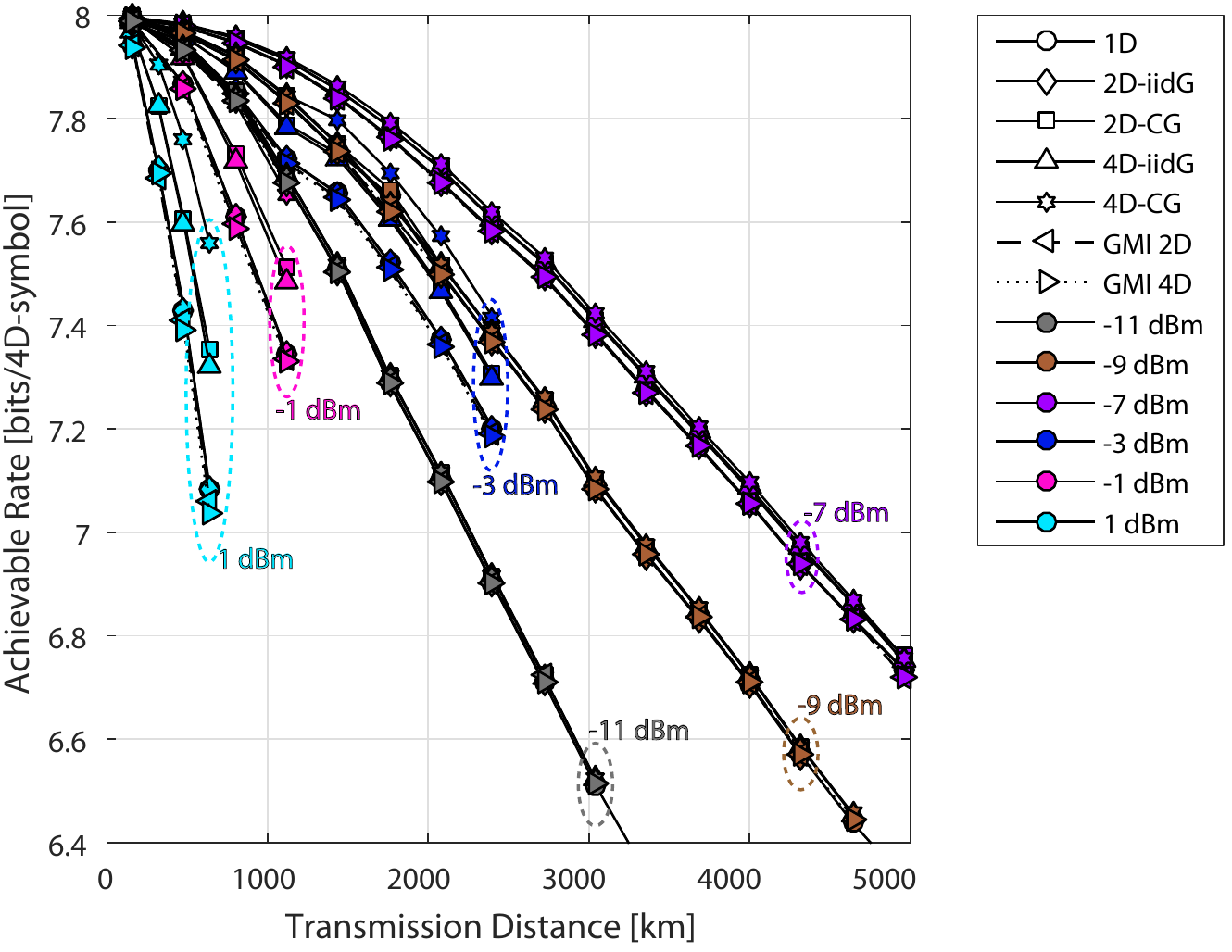}
\caption{WDM transmission of 10 Gbaud PM-16QAM with 15~GHz channel separation, without dispersion compensation.}
\label{PM16QAM_10Gbaud_WDM_15GHz_noDCM}
\end{figure}

\begin{figure}[!t]
\centering
\includegraphics[width=1\columnwidth]{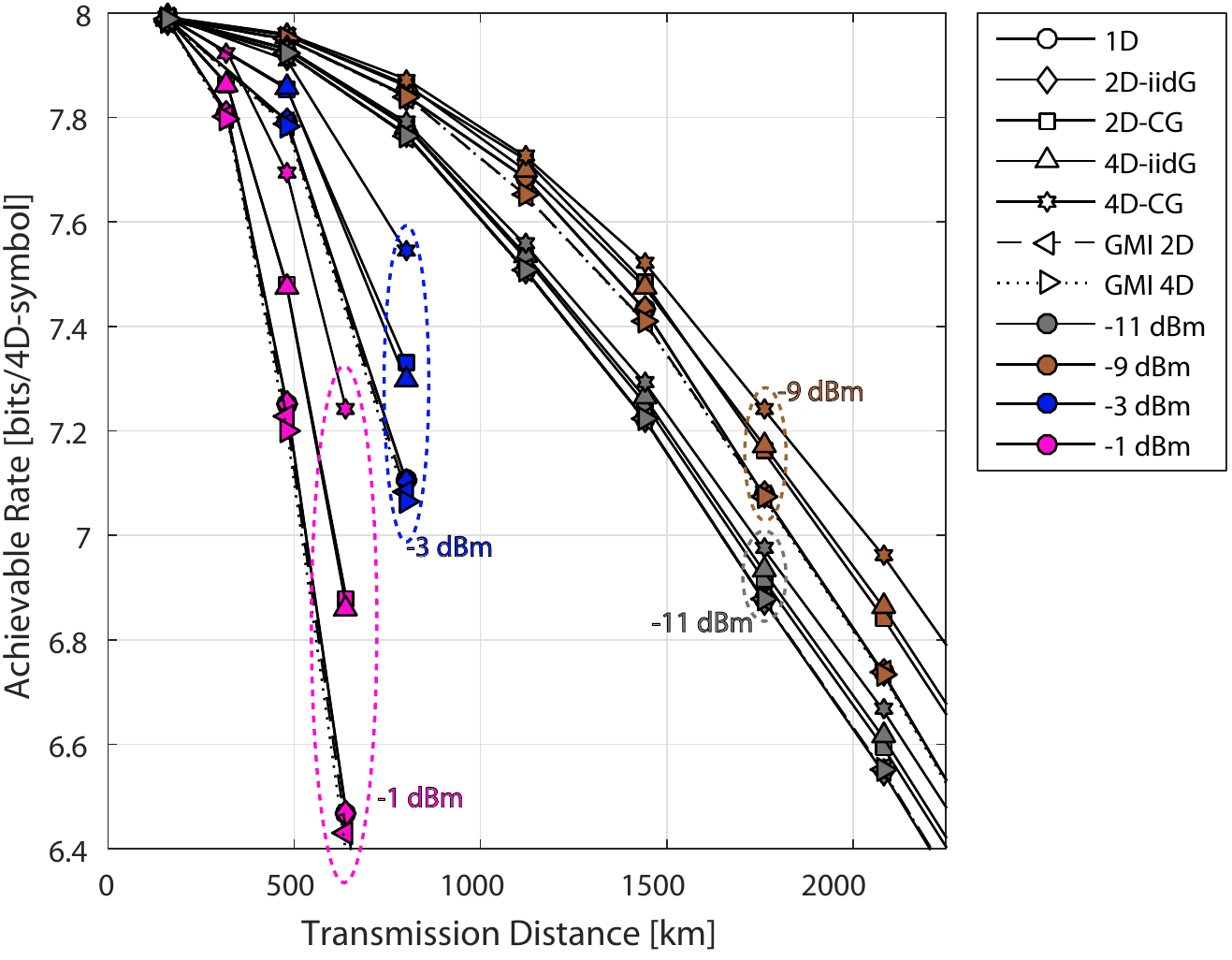}
\caption{WDM transmission of 10 Gbaud PM-16QAM with 15~GHz channel separation, with inline dispersion compensation.} %Note that some launch powers are not plotted to make the figure more clear.}
\label{PM16QAM_10Gbaud_WDM_15GHz_withDCM}
\end{figure}

\subsection{WDM transmission of 10~Gbaud PM-16QAM, without inline dispersion compensation}
In this part, the symbol rate is decreased to 10~Gbaud, while the SE is kept constant by changing the channel spacing to 15~GHz. Shown in Fig.~\ref{PM16QAM_10Gbaud_WDM_15GHz_noDCM} is the achievable rate as a function of transmission distance in this case and for the link without inline dispersion compensation. For the optimal launch power of $-$7~dBm we note, opposed to the 20~Gbaud scenario over the same link, that a difference can be observed between the different estimates. At 4000~km, 2D-CG and 4D-iidG see a gain of roughly 0.03~bit/4D-sym over 2D-iidG. 4D-CG sees the largest gain of 0.04~bit/4D-sym. While these gains are not large, we however note that even if the WDM link is without inline dispersion compensation, if the symbol rate is low enough, the memoryless statistic of the received signal is not perfectly described by 2D-iidG distributions.

\subsection{WDM transmission of 10~Gbaud PM-16QAM, with inline dispersion compensation}
\begin{table*}[t]
\centering
 \renewcommand{\arraystretch}{1.2}
  \begin{threeparttable}[b]
   \caption{Transmission distance difference in percentage compared to the commonly used 2D-iidG distribution at 7~bit/4D-symb.}
   \label{Table2}
   \begin{tabularx}{1.5\columnwidth}{|X | c | c | c | c | c | c |}
\hline
{System Setup}                                      & {1D-iidG}    & {2D-CG}     & {4D-iidG}     & {4D-CG}     & {2D-GMI}      & {4D-GMI}       \\ \hline
WDM 20 Gbaud PM-16QAM, w/o ILDC                     & $\sim0$\%    & $\sim0$\%   & $\sim0$\%     & $\sim0$\%   & $\sim0$\%     & $\sim0$\%      \\ \hline
WDM 20 Gbaud PM-16QAM, w ILDC                       & $\sim0$\%    & 1.8\%       & 2.2\%         & 3.9\%       & $\sim0$\%     & $\sim0$\%      \\ \hhline{|=|=|=|=|=|=|=|}

SC 20 Gbaud PM-16QAM, w/o ILDC\tnote{$\dagger$}     & $\sim0$\%    & 0.5\%       & 0.9\%         & 1.6\%       & $\sim0$\%     & $\sim0$\%      \\ \hline
SC 20 Gbaud PM-16QAM, w ILDC                        & 0.2\%        & 2.5\%       & 2.8\%         & 5.1\%       & $\sim0$\%     & $\sim0$\%      \\ \hhline{|=|=|=|=|=|=|=|}

WDM 10 Gbaud PM-16QAM, w ILDC                       & $\sim0$\%    & 4.6\%       & 5.9\%         & 11.2\%       & $\sim0$\%     & $\sim0$\%      \\ \hline
WDM 10 Gbaud PM-16QAM, w/o ILDC                     & $\sim0$\%    & 1.9\%       & 2.3\%         & 2.9\%       & $\sim0$\%     & $\sim0$\%       \\ \hline

 \end{tabularx}
     \begin{tablenotes}
  %\item [$\S$]  Degrees of freedom. The calculated number is for PM-16QAM.
  \item [$\dagger$] Measured at 7.35~bit/4D-symb instead of 7.00~bit/4D-symb.
  %\item [$\ddagger$] The total R is the sum of the R in both polarizations.
   \end{tablenotes}
  \end{threeparttable}
\end{table*}
Depicted in Fig.~\ref{PM16QAM_10Gbaud_WDM_15GHz_withDCM} is the achievable rate for WDM transmission with 15~GHz channel spacing for 10~Gbaud PM-16QAM with inline dispersion compensation. Some measured launch powers are not plotted for clarity reasons. Note that this case has the same SE as the 20~Gbaud case with 30~GHz channel spacing (Fig.~\ref{Fig_PM16QAM_20Gbaud_30GHz_withDCM}). For the optimal launch power there is a clear difference between the different estimates. At 1760~km, 2D-CG sees a gain of 0.07~bit/4D-sym and 4D-iidG have a gain of 0.10~bit/4D-sym. 4D-CG has the largest gain of 0.17~bit/4D-sym.

\section{Discussion}\label{section:discussion}
First and foremost, we note that for the most realistic scenario, i.e. 10 or 20~Gbaud WDM transmission without any inline dispersion compensation, it is a good approximation to assume 2D iid Gaussian noise distributions, which is most commonly done in \rl{decoders}. This is true also for single-channel transmission over links without inline dispersion compensation. Presented in Table~\ref{Table2} is the approximate gain in transmission distance over a 2D-iidG \rl{channel distribution assumption}, for the optimal launch power at 7~bit/4D-sym for all the different transmission scenarios considered in this paper. We note that for the scenarios considered here, there is never any significant difference between the 2D-iidG, 2D-GMI, 4D-GMI, and 1D-iidG channel assumptions.

\begin{figure}[!t]
\centering
\includegraphics[width=1\columnwidth]{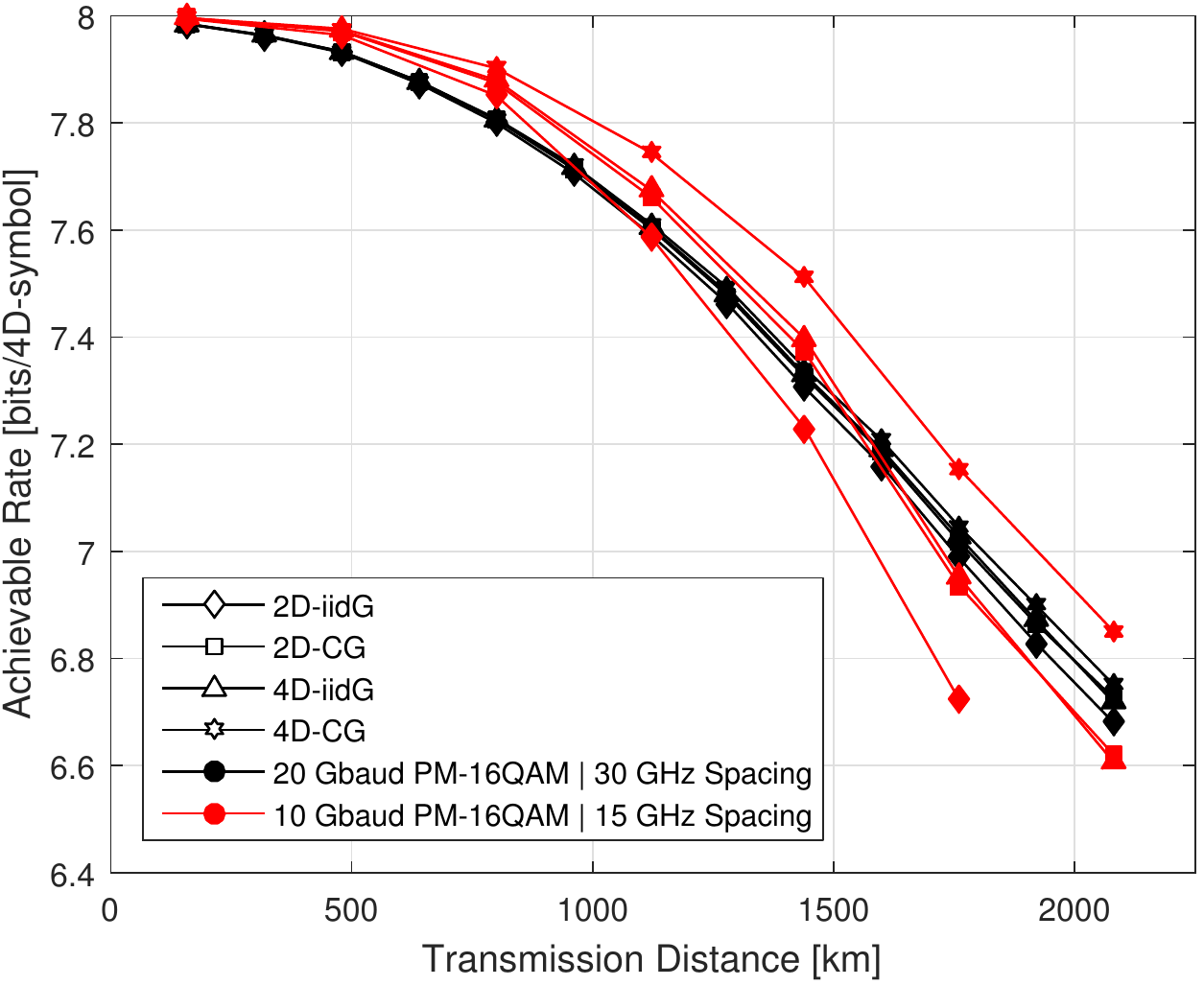}
\caption{Comparison between the achievable rates for two WDM scenarios with the same SE, namely, 20~Gbaud PM-16QAM on 30~GHz spacing and 10~Gbaud PM-16QAM on 15~GHz spacing for the link with inline dispersion compensation.}
\label{Comparison_20and10Gbaud_WDM_withDCM}
\end{figure}

An interesting comparison is the WDM transmission of 20~Gbaud PM-16QAM with 30~GHz channel separation and 10~Gbaud PM-16QAM with 15~GHz spacing, as these two cases have the same spectral efficiency. Shown in Fig.~\ref{Comparison_20and10Gbaud_WDM_withDCM} are these two cases for the 2D-iidG, 2D-CG, 4D-iidG, and 4D-CG estimates at the optimal launch power. We first note that for the 2D-iidG estimate, the longest transmission distance is achieved by the 20~Gbaud case for achievable rates below 7.7~bit/4D-symb. We also note that the difference between the estimates is small in the 20~Gbaud case while for the 10~Gbaud case, a large difference is observed. The 2D-CG and 4D-iidG estimates for 10~Gbaud see roughly the same achievable rates as for the 20~Gbaud case. \rl{For high achievable rates, the 10~Gbaud system achieves a longer transmission distance compared to the 20~Gbaud case for any estimate. This is most likely due to the fact that the number of channels are kept the same, i.e. the spectral width that the 10~Gbaud case occupies is half of that of the 20~Gbaud case. More} notably though is that the 4D-CG estimate for 10~Gbaud achieves the longest transmission distance for any achievable rate.
\begin{figure}[t]
\centering
\includegraphics[width=1\columnwidth]{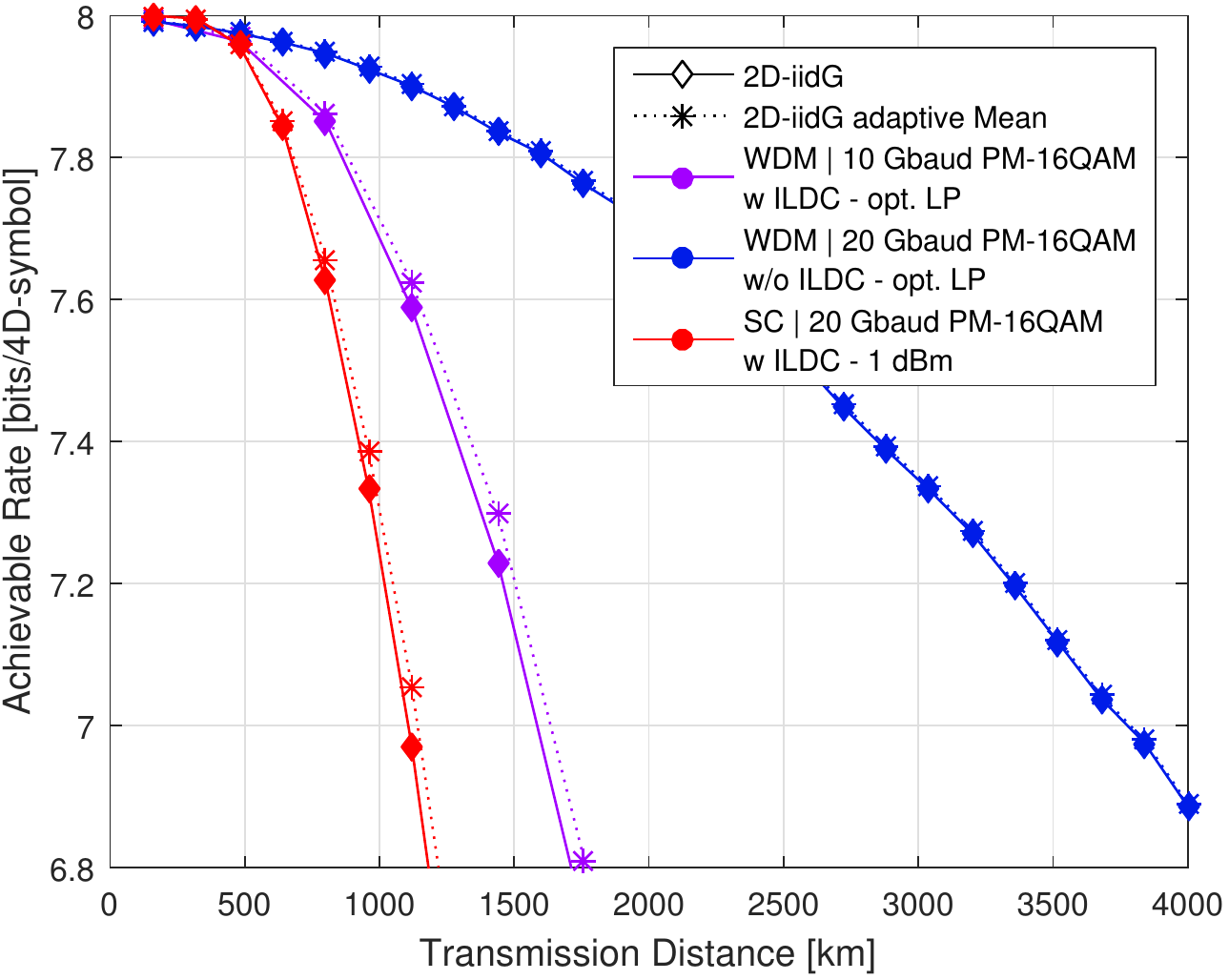}
\caption{Comparison between 2D-iidG estimate with and without adaptive mean values of the constellation points.}
\label{Comparison_2DiidG_wAndWo_adaptMean}
\end{figure}
We conclude that for the system with inline dispersion compensation, there seems to be different optimal symbol rates for a fixed SE, depending on which distribution is assumed in the decoder. \rl{This is a similar finding to that of \cite{Poggiolini15,Carbo15} where it is shown that the nonlinear interference can be reduced by dividing a signal into lower symbol rate subcarriers.} However, a more detailed study on several different symbol rates is required to fully understand this effect. One possible explanation for the higher achievable rate for the 10~Gbaud case is that all the estimates in this paper are memoryless and in the 20~Gbaud case the dispersion increases the nonlinear memory of the channel more rapidly than in the 10~Gbaud case, hence there is more loss of information for the 20~Gbaud signal. We also investigated this for a system without inline compensation and no such effect is then observed.

The baseline measure in this paper is 2D-iidG with non-adaptive means of the constellation points. We note that for a nonlinear channel, some of the gain for the more complex estimates are indeed from this adaptation and not only due to the more sophisticated variance estimates. In Fig.~\ref{Comparison_2DiidG_wAndWo_adaptMean}, 2D-iidG with fixed means is compared to the same case but with adaptive means for three different transmission scenarios. We note that for the optimal launch power for WDM transmission of 20~Gbaud PM-16QAM without inline dispersion compensation there is no distinctive difference, which is expected as we saw no significant difference for any estimate for this case (Fig.~\ref{Fig_PM16QAM_20Gbaud_30GHz_noDCM}). For the optimal launch power of WDM transmission of 10~Gbaud PM-16QAM in the link with inline dispersion compensation, a small gain in the achievable rate can be observed. The same is true for the high launch power of 1~dBm for single-channel transmission of 20~Gbaud PM-16QAM over the same link.

\begin{figure}[!t]
\centering
\includegraphics[width=1\columnwidth]{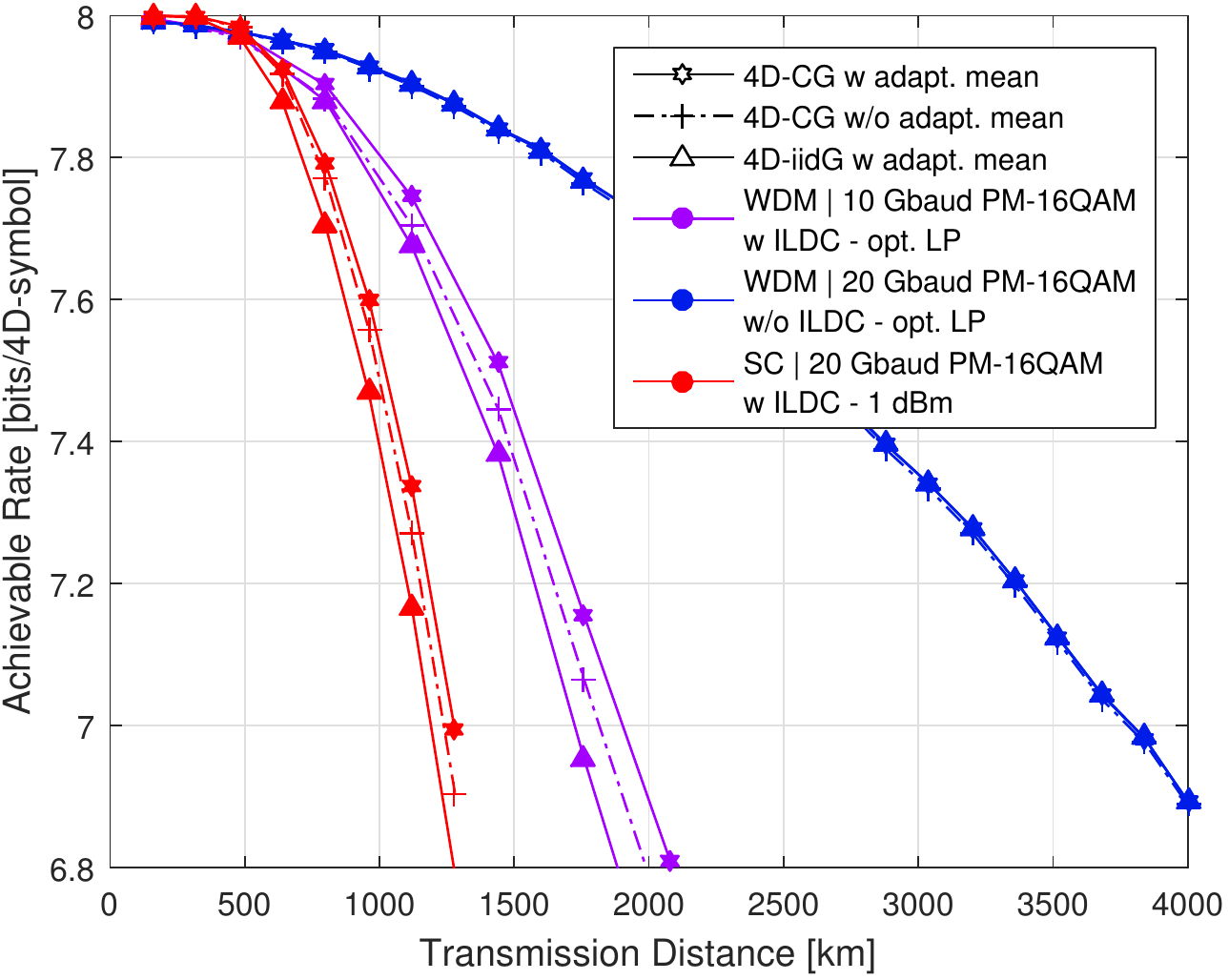}
\caption{Comparison between 4D-CG with and without adaptive mean values for the 4D-constellation points and 4-iidG with adaptive means.}
\label{Comparison_4DCG_wAndWo_adaptMean}
\end{figure}

In Fig.~\ref{Comparison_4DCG_wAndWo_adaptMean} we  compare the best performing estimate, 4D-CG, with and without adaptive mean values of the received constellation points for the same cases as for 2D-iidG in Fig.~\ref{Comparison_2DiidG_wAndWo_adaptMean}. We also added 4D-iidG with adaptive means in this comparison. Again, there is no difference for the WDM transmission without inline dispersion compensation. However, for the two cases with inline dispersion compensation we can indeed conclude that some of the gain for this format comes from the adaptation of the constellation points. However, if we compare to the 4D-iidG with adaptive means, where each 4D-constellation point is considered to have the same variance in all dimensions, we can see that the achievable rate is considerable lower than the 4D-CG case without adaptive means. Hence, we can draw the conclusions that the gain seen for the 4D estimates is dependent on both the fact that the means are adapted in a 4D space and on the 4D estimates of the variances.

\begin{figure}[!t]
\centering
\includegraphics[width=1\columnwidth]{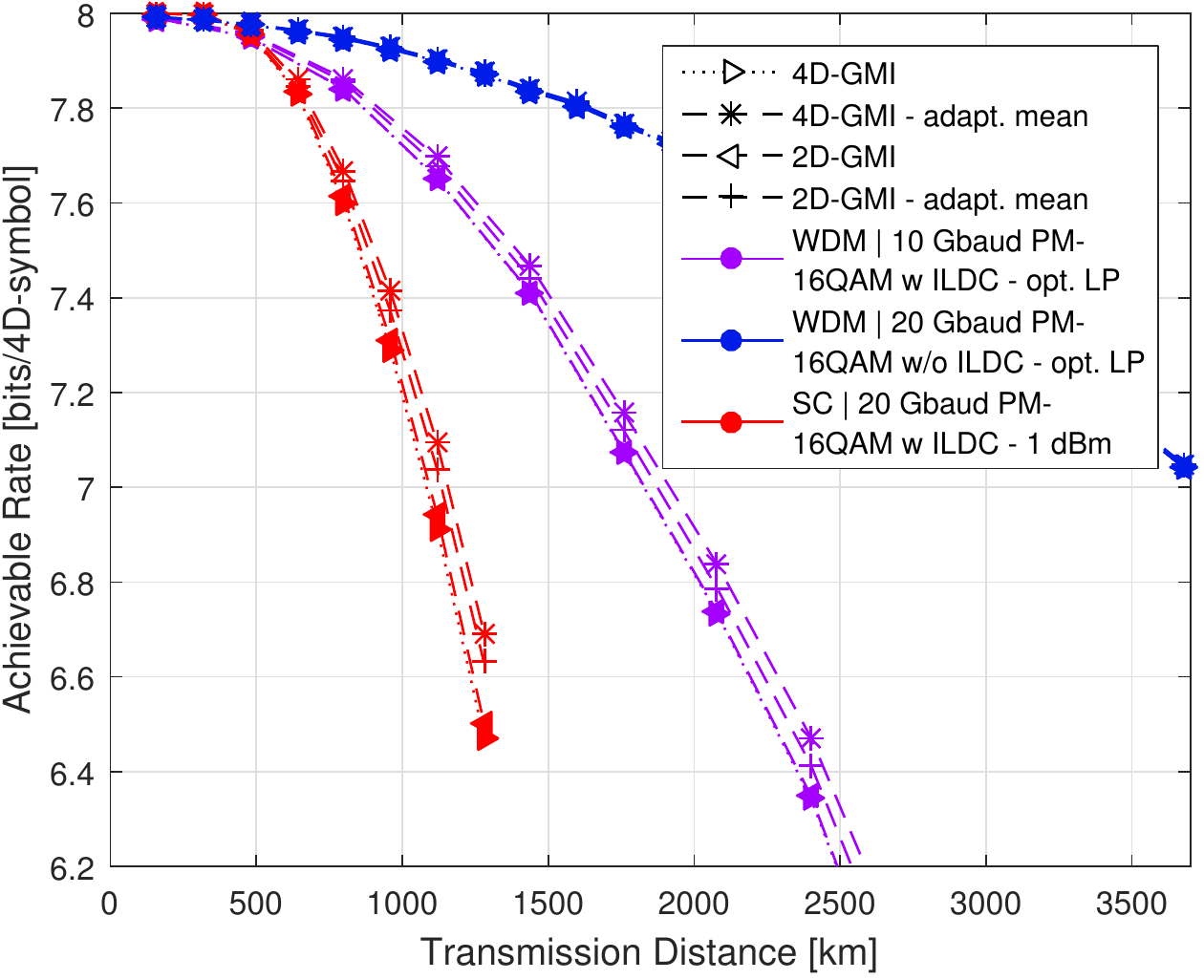}
\caption{Achievable rates for 4D and 2D GMI, with and without adaptive mean values of the received constellation.}
\label{Comparison_GMI_adaptMean}
\end{figure}

Another interesting comparison regarding the adaptive mean of the received constellation points concerns the GMI estimates. In this paper we have compared different MI estimates to 2D and 4D GMI estimates, where we use a fixed received constellation. The reason for this is that it corresponds to the most conventional decoder structures. However, in Fig.~\ref{Comparison_GMI_adaptMean} we compare the 2D and 4D GMI estimates to the case where we allow adaptive estimation of the received constellation points for WDM transmission of 20~Gbaud PM-16QAM without inline dispersion compensation at the optimal launch power, WDM transmission of 10~Gbaud PM-16QAM with inline dispersion compensation at the optimal launch power, and single-channel transmission of 20~Gbaud PM-16QAM at a higher-than-optimal launch power of $-$1~dBm. For WDM transmission of 20~Gbaud PM-16QAM without inline dispersion compensation, no difference is seen between the compared cases. For the systems without inline dispersion compensation with either 10 Gbaud WDM transmission or 20 Gbaud single-channel transmission, we note that there is no significant difference between 2D and 4D GMI without adaptive mean values. However, the 4D-GMI estimate with adaptive mean values sees a clear gain over the other GMI estimates for these two cases.  We also note that the 2D-GMI with adaptive means sees an intermediate gain between not estimating the mean values and estimating the mean values in four dimensions. At 2400~km, 4D-GMI with adaptive means sees a gain of 0.13~bit/4D-sym over 2D-GMI without adaptive means. This shows that for the bit-wise decoder, for certain transmission systems such as the WDM transmission of 10~Gbaud PM-16QAM over transmission link with inline dispersion compensation, there are possible gains by designing decoders that can estimate the mean values of the received constellation points in four dimensions. We note that this should be a minor tweak to existing 4D decoders \cite{Alvarado4DGMI,Batshon4DLPDC}.

\section{Conclusions}
We have experimentally investigated the achievable rate using GMI and MI with different assumptions on the channel distribution for single-channel and WDM transmission of 10~Gbaud and 20~Gbaud PM-16QAM signals for transmission links with and without inline dispersion compensation. We have shown that for most practical scenarios, assuming that the transmission channel has independent and identically distributed Gaussian distribution of the noise in each dimension is a good approximation. In other words, for a practical system, decoders that are assuming fixed constellation means and the same variance for all constellation points have no significant penalty over decoders using more sophisticated distributions. However, for systems with inline dispersion compensation, we show that there is gain in using 4D distributions, most notably using 4D correlated Gaussian distributions with adaptive mean values which shows a small but significant gain even at the optimal launch power. We also show that for all cases, the difference between GMI and MI estimates using 2D-iidG or 1D-iidG distributions is negligible. For the more extreme cases, for instance single-channel transmission with high launch powers, the assumption that the channel is Gaussian with the same variance in each dimension is no longer valid and large gains are seen by assuming 4D correlated Gaussian distributions.

\section*{Acknowledgments}
The authors would like to acknowledge fruitful discussion with Pontus Johannisson and Abel Lorences-Riesgo from Chalmers University of Technology.

\bibliographystyle{IEEEtran}
\bibliography{IEEEabrv,references}
%\begin{IEEEbiographynophoto}{Authors biographies not included at authors request due to space constraints.}
%\end{IEEEbiographynophoto}

\end{document}